# On the Use of Computer Programs as Money

ROSS D. KING, University of Manchester

Money is a technology for promoting economic prosperity. Over history money has become increasingly abstract, it used to be hardware, gold coins and the like, now it is mostly software, data structures located in banks. This data-money is passive and requires computer programs for manipulation, these programs create money, add/remove money from accounts, transfer money in sales transactions, protect access to money, etc. Here I propose the logical conclusion of the abstraction of money: to use as money the most general form of information - *computer programs*. The key advantage that using programs for money (program-money) adds to the technology of money is agency. Program-money is active and thereby can fully participate in economics as economic agents. I describe the three basic technologies required to implement program-money: computational languages/logics to unambiguously describe program-money's actions and interactions; computational cryptography to ensure that only the correct actions and interactions are performed; and a distributed computational environment in which the money can execute. I demonstrate that most of the technology for program-money has already been developed. The adoption of program-money transfers responsibility from human economic agents to money itself and has great potential economic advantages over the current passive form of money. For example in microeconomics, adding agency to money will simplify the exchange of ownership, ensure money is only used legally, automate the negotiation and forming of contracts, etc. Similar advantages occur in macroeconomics, where for example the control of the money supply could be transferred from central banks to money. It is also possible to envisage money that is not owned by any external human agent or corporation. One motivation for this is to force economic systems to behave more rationally and/or more like a specific economic theory, thereby increasing the success of economic forecasting.

・ Applied Computing → Electronic commerce  • Applied Computing → Law, social and behavioral sciences → Economics

## 1. INTRODUCTION

The concept of money is deep and subtle. To quote the Encyclopaedia Britannica: "The subject of money has fascinated wise men from the time of Aristotle to the present day because it is so full of mystery and paradox. The piece of paper labelled 1 dollar, 10 euros, 100 yuan, or 1,000 yen is little different, as paper, from a piece of the same size torn from a newspaper or magazine, yet it will enable its bearer to command some measure of food, drink, clothing, and the remaining goods of life while the other is fit only to light the fire. Whence the difference?" (Friedman, & Meltzer, 2014).

The view presented in this paper is that money is a technology for promoting economic prosperity, and like any technology, it can, and should, be improved: "The special commodity or medium that we call money has a long and interesting history. And since we are so dependent on our use of it and so much controlled and motivated by the wish to have more of it or not to lose what we have we may become irrational in thinking about it and fail to be able to reason about it like about a technology, such as radio, to be used more or less efficiently" (Nash, 2005).

The technology of money has qualitatively advanced several times in history, and with each advance there has been a major increase in economic activity and prosperity (Davies, 1994). Each advance has involved money becoming more abstract. Here I propose the logical conclusion of this process: to use as money the most general form of information - *computer programs*. This proposal is more general than the use of crypto-currencies, such as bitcoin, where the main innovation is the decentralisation of the fiat (Nakamoto, 2009; ECB, 2012; Ali et al, 2014a; Ali et al, 2014b). It also differs in kind from the concept of *machina economicus*, the participation of AI agents in economics (Parkes & Wellman, 2015).

## 2. THE TECHNOLOGY OF MONEY

The essence of money is the transfer of trust between economic agents (Kocherlakota, 1996; Nash, 2005). This involves money functioning (Davies, 1994) as:

- A common measure of value: "money is the 'lubrication' which enables the efficient 'transfer of utility' between economic games" (Nash, 2005). In a market with different goods if there did not exist a common measure of value then there would need to be a quadratic number of exchange rates between goods.
- A medium of exchange. This enables the separation of buying from selling, thereby enabling trade without the "double coincidence" required for barter.
- An asset. Money serves as a store of purchasing power between sales and purchases. Money is by definition the most "liquid" of all assets.
- A causative factor in the economy. Many of the debates in macroeconomics revolve around the causative role of money.

*2.1 Hardware as Money*

The invention of money was one of the most important in human history (Davies, 1994). It undoubtedly transformed the economy of the time (the introduction of money into a barter economy has repeatedly been observed to cause a huge increase in economic activity (Davies, 1994; Friedman, & Meltzer, 2014), and the ramifications of the invention are still unfolding. The first money technology was "commodity money", and money's value comes directly from the commodity from which it was made. Examples of such commodities are barley, cattle, bronze knives, gold, silver, etc. Note that although there is a distinction between commodities of direct utility such as cattle (you can eat them), and commodities of less direct utility such as gold, the monetary role of a commodity often dominates, e.g. in societies that use cattle as money the condition of the animal does not matter, just as an old dollar note has the same value as a new one (Davies, 1994).

One of the earliest recorded forms of commodity money was the Shekel, which originated in Sumer around five thousand years ago as a certain measure (weight) of barley. Barley is far from an ideal monetary technology because of its high volume to value ratio, and the danger of spoilage. However, it is noteworthy that Ptolemaic Egypt solved many of these problems by setting up a common grain storage/bank, and using giros (pointers) to transfer money between accounts – an illustration of the sophistication of even early monetary technologies.

Precious metals, especially gold and silver, offer clear advantages as a technology for commodity money: they do not easily suffer environmental damage, and can be easily stored and exchanged. These advantages have led to their use as money in many societies. The key problem with precious metals as money is ensuring weight and purity. This was solved by the invention (~700 B.C.), probably in Lydia, of the technology of coin money. In coin money a mint places a hard to forge mark on coins that serves as a guarantee of the weight and purity of the metal. Coins also enable payment to be by count, rather than weight, which greatly simplifies commerce. The utility of coin technology has long been recognised: "The various necessities of life are not easily carried about, and hence man agreed to employ in their dealings with each other something which was intrinsically useful and easily applicable to the purposes of life, for example, iron, silver, and the like. Of this the value was at first measured by size and weight, but in process of time they put a stamp upon it, to save the trouble of weighing and to mark the value. (Aristotle)" (Friedman, & Meltzer, 2014). The introduction of coin money caused an economic boom (Davies, 1994). Governments typically impose a monopoly on the creation of coin money, and often impose a fee (tax) on making coins - seigniorage. They are also frequently break the implied contract of guaranteed weight and purity of coins, i.e. they debase the currency.

The next stage in the development of the technology of money was the development of "representative money". In this form of money a certificate or token can be exchanged for the underlying commodity. This eases the difficulties of exchanging commodities. The Song Dynasty in China invented the technology of paper money in around the 10th century AD. This money was "fiduciary": promises to pay specified amounts of gold and silver. The technology of such fiduciary money greatly aided the growth of world trade in the modern era

(Davies, 1994). The early 20th century was the great era of fiduciary money: gold coins circulated in most of the world, and paper money was convertible on demand into gold coins at an official price. The result was in effect a single world currency, and a huge increase in trade (Davies, 1994).

The technology of representative money led to the development of "fiat" money. This is money that has value only because of decree (usually from a Government). "Fiat Money is Representative (or token) Money (i.e. something the intrinsic value of the material substance of which is divorced from its monetary face value) - now generally made of paper except in the case of small denominations - which is created and issued by the State, but is not convertible by law into anything other than itself, and has no fixed value in terms of an objective standard." (Keynes, J.M., 1930). The use of fiat money was originally highly controversial, and forced on nations by wars and economic depression. Now all major currencies are fiat.

*2.2. Data as Money*

In 1930, when Keynes wrote that fiat money was "generally made of paper", most money was already not in the form of paper notes, but in paper records in banks (Davies, 1994; Friedman & Meltzer, 2014). Money had explicitly become data (Kocherlakota, 1996). Now, with the dominance of computers in information processing, only a very small, and diminishing, proportion of money is recorded on paper (Ali & Barrdear, 2014). The vast majority of data-money is electronic data in banks: data-money. This data-money records a series claims, ('IOUs') held by central banks, commercial banks, and their customers. Most of this money is created through bank lending (Davies, 1994; Friedman & Meltzer, 2014; McLeay, et al., 2014).

The following properties of electronic data enable it to serve as money:

- It is *a numeraire*. Goods and services may be valued on a common scale of measurement – digital numbers. This enables data-money to be arbitrarily divisible, e.g. to facilitate micro-payments.
- It may be used as a *medium of exchange*. Data-money may be exchanged for goods and services very easily (liquidity). It is easily transported, and with modern telecommunications it can be sent around the World for almost no cost at close to the speed of light.
- It may be used as an *asset*. Data-money is easily and cheaply stored, and as it is digital, it is potentially stable over the very long-term.
- It is *hard to counterfeit*. Cryptography and other security devices are used to restrict access to the computers that hold the data structures that represent money. Similar techniques are used to ensure data-money is only generated by Governments/banks.
- It is *identifiable/anonymous*. Data-money may be anonymous "bearer instruments", or have an identifiable owner.

These properties have ensured that data-money has dominated economics for decades.

The abstraction of money from hardware to data has had the effect of blurring its definition, with the result that there are different "measures" of money, e.g. M1 a narrow measure of money (medium of exchange), M2, a broader measure (asset money) (Anna J. Schwartz, Money supply, The concise encyclopedia of economics). The main difference between these measures is which categories of commercial and central bank data are "money": demand deposits, savings deposits, central bank credit, assets with zero maturity, etc.

*2.3. Novel Forms of Data-Money*

The vast majority of data money is denominated in national currencies, and held in banks. However, recently, a number of novel forms of data-money have appeared. When data-money is withdrawn and stored in a physical device it is termed 'Electronic-money' (ECB, 2012; Ali et al, 2014a; Ali et al, 2014b). The money withdrawn normally becomes a claim on the issuer. A typical example of electronic-money is M-Pesa, a popular mobile phone service in Kenya that enables access to financial services, including payments, to users (ECB, 2012). M-Pesa has helped spread banking services to the many poor Kenyans who don't have bank accounts.

A virtual-currency is a form of electronic-money used and accepted among members of a virtual community (ECB, 2012; EBA, 2014). Virtual currencies are typically issued and controlled by the developers of the virtual community, they are not normally regulated by financial authorities, and their unit of account has no direct counterpart with a standard currency. A typical example is World of Warcraft (WoW) Gold. This virtual-currency is used as a means of exchange in he online role-playing game WoW, for example to buy equipment to reach higher levels, etc. The buying and selling WoW Gold in the real world is strictly forbidden under the terms and conditions of the game, but there is a thriving black market (ECB, 2012).

A Cryptocurrency is a form of electronic-money where the fiat and technology is decentralised (Nakamoto, 2009; ECB, 2012; Ali et al, 2014a; Ali et al, 2014b; Antonopoulos, 2014). In contrast to standard data-money, cryptocurrencies are not a claim on anybody/anything. The legal status of cryptocurrencies is currently under flux: in some cases they are treated as money, in others as property ('virtual asset'), and in many countries they are illegal (ECB, 2012; EBA 2014).

The original, and most successful, crypto-currency is that of bitcoins, created in 2009 by the mysterious group or individual known as Satoshi Nakamoto (Nakamoto, 2009; O'Hagan, 2016). It currently (2016) has several million users, and the current value of bitcoins in circulation is several billion $US. The bitcoin system is a sophisticated combination of public-key cryptography (to identify accounts), peer-to-peer public-ledgers that records transactions (the data-money), and the use of a computation proof-of-work (POW) system (based on game-theory) to time-stamp transactions and thereby decrease the possibility of double-spending. The combined system ensures that distrustful agents can work together to exchange trust (i.e. bitcoins): standard users hold bitcoins and use them as a unit of exchange, and as a speculative asset; and special users, known as 'miners', hold copies of the public-ledge, gather together blocks of transactions, and compete to verify them.

The key design goal in developing the bitcoin system seems to have been to develop a form of money that avoids use of a centralized control (Nakamoto, 2009). The motivation seems to have been political/financial: the first block in bitcoin's block chain (the 'genesis block') includes the text: 'The Times 03/Jan/2009 Chancellor on brink of second bailout for banks', a reference to a newspaper article from that day (Ali et al, 2014a). Proponents of bitcoins argue that the decentralized fiat means that with banks don't have a monopoly on money creation, there is a lower risk of seizure of one's bitcoins by governments (which control the banks), and it is banks can't make unfair profits from transaction costs. The perceived lower risk of bitcoin seizure compared to traditional forms of money has attracted the attention of criminals, and bitcoins have notoriously been used to buy illegal weapons, drugs, etc. (Ali et al, 2014a). The current fees on the bitcoin network tend to be less 1% of the transaction, which compares well with the 2%-4% for traditional online payment systems (EBA (2014) - despite the marginal cost for the bitcoin system being higher (Ali et al, 2014b).

Despite its undoubted success, the bitcoin system has a number of serious shortcomings: lack of true anonymity of transactions, difficulty in scaling, and an environmentally damaging POW scheme. Newer cryptocurrencies such as Monero, Litecoin, Peercoin etc. have been developed to overcome these disadvantages. Bitcoins are often described as 'anonymous', but this is incorrect. A bitcoin is similar to cash in that parties can transfer ownership of the money without disclosing their identities to a third party or to each other. However the transactions to and from a particular bitcoin address can be traced as they are part of the ledger, and as the ledger is public a hostile agent has the opportunity to use techniques such as network analysis to determine the owner of a bitcoin address (Bohannon, 2016). A fundamental weakness of the bitcoin system is that it is unlikely to be able to scale to be deal with a significant proportion of the world's financial transactions. So far the number of global bitcoin transactions has never exceeded 100,000 per day, which compares with ~300 million daily conventional transactions in Europe alone (EBA 2014). One key scaling difficulty is that miners have to maintain a public-ledger containing all previous transactions. This is unlikely to work when used for a significant proportion of the world's financial transactions, and with an increased number of transactions the marginal costs for transactions is likely to rise above those of centralized control systems (Ali et al, 2014b). However, arguably the greatest disadvantage of the bitcoin system is the financial cost/environmental damage of the POW system. This depends on CPU power to ensure the irreversibility of transactions and so avoids double-spending. Specialized software and hardware has been developed to do this, but the bitcoin POW consumes an increasingly large amount of electrical power. This fundamental feature of POW schemes has led to the development of proof of stake (POS) schemes aimed at securing a cryptocurrency network and achieving distributed consensus. POS schemes are based on asking users to prove ownership of a certain amount of currency (stake) in the currency.

Perhaps the most significant aspect of the success of bitcoins and related cryptocurrencies is that it they inspired the development of new software tools for computational financial instruments. These new tools have taken advantages of developments in cryptography, game-theory, artificial intelligence, etc. to widen the range of possible forms of money.

### 3. COMPUTER PROGRAMS AS MONEY

*I propose extending the technology of money from computer data to general computer programs: program-money*. This is the natural end-point of the increasing abstraction of money: from hardware, to data, to programs. There can be no more abstract and general form of money.

Despite the outstanding success of data-money: standard bank data-money, electronic-money, virtual currencies, cryptocurrencies, etc., *data-money is limited and not fully general*. It's greatest limitation is that it is just data, and as such passive. By this I mean that it requires computer programs for manipulation. These programs create money, add/remove money from accounts, transfer money in sales transactions, protect access to money, etc.

The key advantage that program-money adds to the technology of money is agency. Program-money is active and thereby can fully participate in economics as an independent economic agent. (By "agent" I mean the technical computer science sense: "an agent is any entity, embedded in a real or artificial world, that can observe the changing world and perform actions on the world" (Kowalski, 2011)). There are many things that one might wish money to do but which are currently very difficult because money is passive. For example, in microeconomics, adding agency to money could potentially ensure that money is only used lawfully, tax is not evaded, etc. This could be achieved by transferring the responsibility for paying tax from (honest/dishonest) human agents to money itself. Similarly, a possible application in macroeconomics would be the use of computer money to control inflation, or deflation. This could be achieved by transferring the responsibility for money supply to money itself by directly programming it to increase or decrease in quantity over time. It is also possible to envisage money that is not owned by any external human agent or corporation. One motivation for this is to force economic systems to behave more rationally.

The development of program-money requires:

1. Computational languages/logics to unambiguously describe program-money's actions and interactions
2. Computational cryptography to ensure that only the correct actions and interactions are performed.
3. A distributed computational environment in which the money can execute.

I argue that as the basic technology for all these has already developed, there exists no major technological barriers to the use of computer programs as money

I envisage that the primary supplier and developer of program-money will be national Governments - though private program-money currencies similar to existing cryptocurrencies would be also possible. National governments have the best resources to develop program-money: to correctly program its actions, cryptography, and environments. They also have the necessary resources to enforce correct usage, and punish wrongdoing – as they currently do with data-money. Program-money's use by national Governments would also ensure that its advantages are shared by all citizens.

*3.1 Specifying the actions of program-money*

The actions of program money will define its internal computations and communication with other agents. There are a vast potential number of tasks that program-money could be programmed to do: exchange of ownership, obedience to the law, forming binding contracts, negotiations, etc. Individual units of money will have different capabilities, but also include standard competences and interfaces. A large amount of work has already been done on programming such tasks, and I argue that almost all of the computational methods required for program-money have already been developed. For example a large amount of effort has been put into the area of business process languages (Milli et al., 2010).

Just as with existing software for the manipulation of data-money, there is a powerful financial incentive to ensure that program-money works as intended, i.e. its semantics are correct. Ideally the execution of program-money should be provably correct. One approach to ensure this is to utilise some form of computational logic to provide declarative semantics (Kowalski, 2011). Two particularly applicable forms of logic are deontic logic, and transaction logic: deontic logic is concerned with obligation, permission, and related concepts is (Gabbay D); transaction Logic (TR)[3] aims to provide a theoretically-grounded behavioral modeling (A.J. Bonner and M. Kifer, 1995).

All economic transactions should be legal, however human economic agents don't always obey laws. Program-money could be engineered to always obey the law, and thereby severely restrict the ability of human agents to disobey financial laws. To ensure that program money obeys the law it is necessary for both the semantics of the law, and of program-money to be clear. This is feasible as there already exists a large research literature on the use of computational logic to clarify the implementation of the law (e.g. Sergot et al., 1986). "Computational law systems take advantage of the direct, semantic connection between the terms in laws, rules, and regulations, and the model of the digitally-mediated world manipulated by agent actions, enabling a rigorous, formal approach to legal reasoning." (Love & Genesereth, 2005). Although some lawyers have criticized research on computational law system on moral and democratic grounds (e.g. Leith, 1986), these criticisms don't apply to the use of logic to ensure that program-money obeys the law.

It will also be beneficial for program-money to participate in business contracts. Contracts are similar to laws with the main difference being that the parties explicitly agree to them. Business contracts are used to specify the obligations, permissions and prohibitions that the signatories should hold responsible for and to state the actions or penalties that may be taken when any of the stated agreements is not being met (Governatori & Rotolo, 2004). As with computational law, a significant amount of research has already been done on the computational specification and implementation of contracts (e.g. Szabo, 1997; Governatori & Rotolo, 2004; Neal et al., 2003; Miles et al., 2010). The concept of "smart contracts" was developed by Nick Szabo (1997) "Smart contracts combine protocols with user interfaces to formalize and secure relationships over computer networks. Objectives and principles for the design of these systems are derived from legal principles, economic theory, and theories of reliable and secure protocols". The concept emphasizes the goal of integrating the well-established practices of contract law with the design of electronic commerce protocols on the Internet

*3.2 Cryptography for program-money*

The World's current financial system is based on fiat data-money, and therefore depends on computational cryptography for its integrity. Computational cryptography is required to ensure the security of data-money, its availability on demand, the non-repudiation of data-money transactions, etc. Computational cryptography is also essential for crypto-currencies. Program-money will also depend on computational cryptography for its integrity.

Modern cryptography is built on the twin foundations of information theory and complexity theory: a secret is secure either because one's adversaries do not have sufficient information available to break it (e.g. a one-time pad), or because one's adversaries do not have sufficient computational resources to break it (e.g. RSA encryption) (Lewis & Papadimitriou 1981; Goldreich & Wigdesrson, 2008; Katz & Lindel, 2014). Complexity theory is the weaker foundation as it less well understood, one-time pads will always be secure, but the discovery that $P=NP$ would precipitate a financial crisis (Cook, 2000).

Financial cryptography is a large and flourishing branch of cryptography, its basic goal is to: "construct financial systems that maintain their desired functionality (rules, privacy requirements, etc.), even in the face of malicious computational attempts to make them deviate from this functionality" (Goldreich & Wigdesrson, 2008; Katz & Lindel, 2014). Almost all of the financial cryptography methods required for program-money have already been developed for electronic data-money. The two most basic cryptographic requirements are permanency, and secure communications. Permanency ensures that program-money cannot be tampered with by an outside malicious agent, e.g. to change ownership, to change the execution of the code, etc. With tamper-proof code it is possible to both detect if the program has been altered (Collberg, 2002; Arnold, et al., 2003). Possible methods of achieving tamper-proofness are digital fingerprinting, program checking, etc. (Collberg, 2002). It may be advisable for program-money to react to attempted tampering by loosing its value (zeroisation), or by message sending a message to an authority. This self-destruction technique is used to safeguard software used in weapons (Arnold, et al., 2003). The requirement to eliminate interference with the code of program-money is complicated by the possibility that the code is running on an unauthorized machine – "a malicious host attack" (Collberg, 2002). This risk is reduced by computational environment for program-money 0 see below.

As with data-money it is potentially very easy to duplicate and spend program-money multiple times: the "double-spending problem". Numerous schemes have been proposed to deal with this. The simplest approach, both technically and politically, is an online central trusted authority that can verify whether more than one copy of signed money exists. Alternatively, much of the excitement surrounding the bitcoin system is in its ledger system that avoids the double-spending problem (Nakamoto, 2009; Ali et al, 2014a; Antonopoulos, 2014).

Program-money will necessarily need to communicate with other economic agents. This communication needs to be secure to avoid it being misled into incorrect behaviour. Securing the communication between agents is the oldest and most basic problem in cryptography, and a vast literature exists on this, e.g. Goldreich & Wigdesrson, 2008; Katz & Lindel, 2014. An important element in secure communications for program-money will be the use of cryptographic signatures, these are cryptographic schemes designed demonstrate the authenticity of a message or document (Katz & Lindel, 2014). A valid digital signature gives the receiver evidence that the message was created by a known sender (authentication), that the sender cannot deny having sent the message (non-repudiation), and that the message was not tampered with in transit (integrity). Cryptographic signatures will be used by program-money in such tasks as transferring ownership between agents. Many protocols for electronic signatures already exist, most of which are based on use of public key encryption (Rivest et al, 1978; Goldreich & Wigdesrson, 2008; Katz & Lindel, 2014): for example a central bank "signs" computer money (the fiat) using its private key, and customers and merchants may verify the signed money using the bank's public key.

Program-money could be engineered to be either identifiable or not. With identifiable money it is possible to trace specific financial interactions throughout the economy. Such identification is easily implemented by the program-money reporting each time it is involved in a transaction. Most existing data fiat data money is identifiable. Being identifiable will simplify the implementation of program-money, and helps ensure laws are obeyed, however it also raises privacy concerns. Some forms of electronic-money are non-identifiable, for example through the use of blind signatures, where the content of the message is hidden (blinded) before signature (Chaum, 1983). A large part of the attraction of bitcoins has been based on the belief that it is non-identifiable. However, as every transaction is recorded, users of bitcoins have to use extra cryptographic techniques to avoid identification (Bohannon, 2016). Several newer forms of cryptocurrency, such as Monero, have attempted to improve on bitcoins by being less identifiable. Governments generally object to non-identifiable money on the grounds that it can easily be used for illegal transaction. As program-money has access to the transactions it is involved in, and can reason about them, it would be possible, unlike with existing non-identifiable electronic money, to implement non-identifiable money that rejects transactions that are illegal.

*3.3. An Environment for the Execution of Program-Money*

The essence of money is the transfer of trust between agents in economic interactions. With program-money these interactions will take place within an especially designed computational environment, and the success of program-money will depend in large part on the sound design of this environment. The application of computational game theory will be required in the design of this environment to ensure that the desired behaviour of the agents is in line with their economic advantage (Nisan, et al. 2007). The Internet will provide the infrastructure to interconnect the hardware running the program-money. Cryptography will ensure that the individual instantiations of program-money are able to communicate safely with each other.

Program-money will of course be executed on physical computers. Different economic agents will own these computers: customers, shops, banks, central banks, etc. To reduce the probability of physical tampering with program-money, this hardware could be required to be secure through the use of secure cryptoprocessors (Waksmann & Sethumandhavan, 2010).

For some application it will be necessary for program-money to monitor its physical location. Again a substantial amount of research has already been done in this area: with the focus on proving the location of mobile devices. Location-proof cryptographic schemes have been proposed using secure GPSs (Denning & MacDoran, 1996), and wireless networks (Lenders et al 2009; Saroiu & Wolman, 2009). If the location-proof system determines that program-money is moved to a non-permitted location it could react in a similar way to attempts to detected attempts to tamper with the program by for example loosing its value (zeroisation), or by message sending a message to an authority. Conversely, program-money could equally well be programmed to hide its geographical location (Jiang, et al 2007)

*3.4. Negotiations Between Program-Money*

Before taking part in a financial transaction it may be necessary for program-money to negotiate with other economic agents, especially other instantiations of program-money. Negotiations between economic agents is known to improve the efficiency of markets, so the ability of program-money's to negotiate will extend the efficiency of negotiation-based markets to many more areas (Muthoo, 1999). Again, little or no new technology would be required for program-money to take part in negotiations. Cooperative game theory applies to negotiations and bargaining, and a sophisticated computational theory of negotiations and interactions between economic agents has been developed (Parkes, & Wellman, 2015). (In controlled experiments computers can outcompete humans in financial games, e.g. double auction markets (standard stock exchanges) (Das et al, 2001)). There exists extensive practical experience of computational agents being involved in financial negotiations. For example, in algorithmic trading programs automatically make trading decisions in stock markets, currency markets, etc., and such programs increasingly dominate financial markets (Cartea, et al., 2015). Most types of algorithmic trading program are based on high-frequency trading, where the speed of computational programs is used to exploit small (often tiny) inefficiencies in markets before rival human and computational agents can. (In 2010 algorithmic trading they were implicated in the infamous "flash crash", in which the Dow Jones Industrial average first plunged ~9%, and then recovered its most of its value within minutes.) Closely related to algorithmic-trading are bidding agents (Wellman, et al. 2007; Nisan, et al. 2007). One important application area for bidding agents are sponsored search auctions (Nisan, et al. 2007). These are sophisticated auctions for advertising space on search engines such as Google, Bing, etc. based on search terms. Bidding agents are essential for these giant markets.

**3. APPLICATIONS**

*4.1. Greater Simplicity*

Program-money also has the potential to ease the administrative burden of following rules and laws associated with money, e.g. tax laws. If I go into a shop and buy a cake, and I'm in a country where there is a sales-tax on cakes, then the vendor is responsible for paying to the Government a fraction of the price of the cake. With program-money it will be possible to shift responsibility for paying the tax from the vendor to the money itself: the money in the customer's wallet would be informed that it owner wishes to buy a cake, the money would then transfer ownership of part of itself (the value of the cake) to the vendor's wallet. The program-money in the vendor's wallet could then check the authenticity of the incoming money using similar cryptography techniques to those currently used for data-money. Similarly, the customer's money could check that the money it is paying will join legitimate money. The money in the vendor's wallet then necessarily recognises that sales tax is due, and therefore transfer ownership of part of itself, the value of the sales tax to the tax-authority. Depending on the protocol the money could recognise the need for sales tax by consulting some external authoritative law server, or through its internal knowledge of the law. Similar processes could implement a "use tax" where the customer's wallet is responsible to pay the tax, or VAT - where there is a chain of responsibility. In this way the use of program-money would transfer the administrative burden of taxation from the vendor/buyer to the money itself.

*4.2. Ensuring Obedience to the Law*

Program-money has the potential to help ensure that money is only used lawfully, and that due taxes are paid. For example, it will be possible to ensure that money is not used to buy anything illegal without a licence. This could be implemented by the money itself checking the details of transactions it is being requested to be involved in, and refusing to take part in unethical purchases. It might be possible to fool the money into committing to a false transaction, but this would be much more difficult that with data-money.

Many Governments require annual declaration of assets, and program-money could be programmed to destroy itself if it were unable to annually contact its Government. Similarly, many Governments impose constraints/taxes on the movement of money across borders, program-money may need to be able determine its current spatial and/or temporal location. Program-money could also make it much harder to illegally hide money abroad in a tax-haven such as Panama. It could for example be programmed to destroy itself if it were unable to confirm that it was executing on a computer in its home country.

*4.3. Greater Control of Your Money*

Almost all current forms of money are fungible, this means that any unit of currency, say a dollar can be exchanged for any another (Davies, 1994; Friedman & Meltzer, 2014; Nangle, 2016). This simplifies economic systems, but also has the disadvantage of lose of control to the owners of money. Program-money has the potential to give the owner of money greater control. For example the owner of money may give another economic agent control over some program-money, but still retain certain restrictions on its use, perhaps not being used to buy anything considered to be unethical. Such restrictions could be implemented by the money itself checking the details of transactions it is being requested to be involved in, and refusing to take part in unethical purchases. Similarly, program-money could make the banking system fairer and money more democratic. It could give the program-money's owner control of the use of their money as collateral by a bank. It could, for example, be impossible/require permission to loan it to borrowers of certain types, e.g. arms manufacturers, etc.

*4.3. Delegation*

An owner of program-money may wish to delegate aspects of control of their to the program-money itself. For example they could instruct their program-money when deposited in a bank to monitor advertised bank interest rates, and to autonomously decide to move location to obtain a better interest rate. Such economic decision making by computational agents on behalf of their owners is already common in financial markets - algorithmic trading. Program-money could democratise this technology and make it generally available. (Analogously, much technology from racing cars makes its may to ordinary cars). The result would be of economic benefit as the markets would be more efficient.

*4.4. Predictive Economics*

Program- money is a politically neutral technology, and it could be used to help implement many different economic/political systems. Macro-economics does not have a good success rate in predicting the future behaviour of economics, for example few central banks or commentators predicted the 2008 crash (Bernanke, 2015). There are two broad approaches to improving the predictive success of economic forecasts. The standard approach is to develop more accurate economic models, as is done in the natural sciences. However, this is very challenging, and possibly intractable, as it is impossible to do repeat controlled experiments in macro-economics; economic relationships are often non-linear, resulting in hard to predict behaviour; economies are so inter-linked that they all need to be predicted at the same time; human economic agents are often irrational (Kahneman, 2012); etc. The more radical approach to improving economic forecasting, one not open to the natural sciences, is to engineer the economic system to better-fit theory. Program-money is able to do this, as adding agency to money could enforce desired forms of causation in the tangled web of financial interactions that constitute economies. For example, program-money could be designed to behave more closely to Keynesian, monetarist, etc. theory. Of course there are limitations what a monetary system can do when most economic agents are humans, but it is reasonable to envisage a system where program-money enforces constraints that make human economic behaviour would more closely resemble theory.

### 4.5. The Money Supply

Money is currently either created directly by fiat by Central banks, or indirectly by Commercial banks (fractional reserve banking) (Davies, 1994; Friedman & Meltzer, 2014; Nangle, 2016). Program-money could provide greater or lesser control over the money supply. For example, program-money could automatically increase or decrease in quantity over time to ensure inflation or deflation. The advantages of such a system over the current one is that that there is greater control and certainty, as the agency belongs to the money not to human bankers. This approach to money supply is similar to that currently used by most cryptocurrencies where the future path of money supply is pre-determined and governed by a protocol that ensures that the eventual total supply will be fixed (Ali et al, 2014b). For example, the bitcoin system is designed to gradually decrease the production of currency, and so places an ultimate cap on the total amount of currency that can ever be in circulation. (It is interesting to consider what would happen if the bitcoin money supply approach was applied to a full economy. The conventional argument is that it would contribute to deflation in the prices of goods and services (and wages), and the inability of the money supply to vary in response to demand would likely cause welfare-destroying volatility in prices and real activity (Ali et al, 2014b). However, economists from the Austrian school would disagree (Hayek, 1976)). In contrast with cryptocurrencies it is possible to envisage inflationary program-money. The simplest example would be a rule in which the money supply were permitted to grow at a constant rate per year, similar to that advocated by Friedman (1969). It is also possible to imagine more general rules that responded to variation in demand, for this a more flexible rule would be required. For example, the growth rate of the program-money supply could be adjusted to respond to transaction volumes in (close to) real time. Alternatively, a decentralized interactive system could be developed to target a fixed exchange rate (Ali et al, 2014b). Program-money could also be time-limited, that is it must be spent by a certain date. One application of such time-limited program-money would be to provide a short-term economic stimulus without the danger of unwanted inflation (Nangle, 2016).

### 4.6. Economic Inequality

Arguably the greatest economic problem facing the world is inequality in wealth. Program-money has the potential to more fairly distribute money so that it has greater utility to more people. The marginal utility of money decreases with quantity (Ramsey, 1931; Stigler, 1950). This has led to the suggestion, that goes back to Daniel Bernoulli, that the utility of money to people is proportional to the log of its quantity. Therefore one possible way program-money could promote economic equality would be by implementing operations based on the utility of money, not its quantity, perhaps based on Bernoulli's log rule. In this case when an owner of program-money obtains more program-money the total calculated by the program-money is calculated based on the utility of money.

## 5. DISCUSSION

*5.1. Objections*

*It is not money.* It could be argued that passivity is a fundamental property of money, and the use of program-money would change the concept of money so radically that it is no longer money. In support of this objection is the fact that all previous technologies for money have been passive. The counter-argument is that the charge 'it is not money' has been made for previous changes in the technology of money (Davies, 1994). Moreover, previous changes in the technology of money have arguably been more profound: moving from commodity money to representative (fiduciary) money involved the radical removal of the intrinsic value of money; and the move to fiat money involved the removal of any involvement of an intrinsically valuable commodity.

*It is not really new.* It could be argued that there is nothing really new in the proposal, and the existence of computer data money already implies the existence of computer programs. This argument has some merits, as many, perhaps most, of the operations of computer program-money could be done by computer programs manipulating data money. However, I argue that there is a fundamental difference of perspective when money has agency. It is a matter of who is pushing whom around.

*It could never work.* Above I have sketched the technology required for programming money, the cryptography required to protect it, and the infrastructure required to run it. This sketch is incomplete, and it is possible that there are insurmountable technological problems that prevent program-money from becoming a reality. For example, the tension between the need for the semantics of the program-money to be clear and understandable (to help ensure correctness), and the need for the same program-money to detect tampering may be insurmountable. Despite this possibility I think there is a strong argument that the basic technology for all aspects of program-money have already been developed: we know how to make the logic of programs explicit and a vast amount of financial software exists, cryptography is already essential in keeping money secure, and the intranet provides the necessary infrastructure

*It is not safe.* It could be argued that there are already a myriad of security concerns with existing data money, and the move to computer programs can only make this worse. It is certainly true that our existing financial system relies on the correct working of computer programs and the Internet, and this makes it vulnerable to security threats to which paper money was immune. For example a bank could face a denial of service attack that makes it difficult or impossible for its customers to spend money, and this could not have happened with previous money technologies. However, it is also true that technology removes security threats, for example with the easy availability of ATM machine there is now less need to carry around large amounts of cash which could easily be stolen. I would therefore argue that program-money could increase the security of money.

*It would erode privacy and freedom.* Depending how a program monetary system was designed it could increase or decrease privacy compared to data-money. Admittedly this is a low base, as it is currently possible to track most money through the economy, even bitcoins. Modern cryptography systems depend on hard functions, and these may not be actually hard to your adversary, or your adversary may have subverted your cryptography code. Such problems are especially acute if the adversary you are concerned about has a disproportionate amount of resources, i.e. a Government.

*It would cause huge economic instability.* Capitalist economies are notoriously prone to instabilities, cycles of growth, and booms and busts. It is possible that such instability could be made worse by program-money.

*5.2. Extensions*

It is possible to envisage money that is not owned by any external human agent or corporation. One motivation for doing would be to promote greater economic rationality, which would benefit society through the better allocation of resources. Traditional economic theory is predicated on the assumption of rational economic behavior (*homo economicus*), i.e. human economic agents acted so as to maximise their own economic benefit (Rittenberg & Trigarthen, 2009). This rationality is often assumed to be computationally bound, i.e. agents have limited computational resources, and must decide how to act within a given amount of time (Simon, 2008). However, even given bounded resources, human economic agents often seem to fail to act as straightforward maximizers of utility, but it is often unclear whether human economic agents are simply stupid, or maximizing a more complicated utility function (Becker & Murphy, 1988). The realization of the importance of human psychology in economic decision making has led to the field of behavioral economics (Kahneman, 2012). This complexity of human economic behavior makes difficult the rational allocation of resources, and makes economic prediction intractable (see above). Many markets are also inefficient because they are not free, as their human agents are influenced by threats, bribes, etc. This results in economic inefficiency, and the theft of resources. Active money could be programmed to act rationally, and to flow to where it will get the best return. It will not be influenced by irrational greed and fear, nor by threats and bribes – which dominate may human markets. Possible also to ensure other types of behaviour, e.g. only investing in companies that meet certain requirements: good labour relations, non-polluting, etc. It is also possible to envisage program-money that learns over time. Such program-money could learn to act more economically efficiently. Such intelligent program-money would approach '*machina economicus*' the full participation of AI agents in economics (Parkes & Wellman, 2015).